\documentclass[twocolumn,showpacs,preprintnumbers,amsmath,amssymb,apsrev,aps,prb]{revtex4}
\usepackage{graphicx}
\usepackage{dcolumn}
\usepackage{bm}
\usepackage{subfigure}
\usepackage[latin1]{inputenc} 
\newcommand{\comment}[1]{}

\usepackage{bm}
\usepackage{color}
\usepackage{subfigure}
\usepackage[latin1]{inputenc} 
\usepackage{ulem}
\begin{document}


\markboth{L. Fang, W.J.T. Bos, L. Shao and J.-P. Bertoglio}{Time-reversibility of Navier-Stokes turbulence}

\title{Time-reversibility of Navier-Stokes turbulence and its implication for subgrid scale models}

\author{Le Fang, Wouter J.T. Bos, Liang Shao and Jean-Pierre Bertoglio}

\vspace{0.2cm}

\affiliation{LMFA-CNRS, Universit\'e de Lyon, Ecole Centrale de Lyon, Universit\'e Lyon 1, INSA Lyon, 69134 Ecully, France}


\begin{abstract}
Among existing subgrid scale models for large-eddy simulation (LES)
some are time-reversible in the sense that the dynamics evolve backwards in time after a transformation $\bm u \rightarrow -\bm u$ at every point in space.
In practice, reversible
subgrid models reduce the numerical stability of the simulations since
the effect of the subgrid scales is no longer strictly
dissipative. This lack of stability constitutes often a criterion to
reject this kind of models. The aim of this paper is to examine
whether time-reversibility can constitute a criterion that a subgrid
model has to fulfill, or has not to. Thereto we investigate by direct
numerical simulation the time-dependence of the kinetic energy of the
resolved scales when the velocity is reversed in all or part of the
lengthscales of the turbulent flow. These results are compared with
results from existing LES subgrid models. It is argued that the
criterion of time-reversibility to assess subgrid models is incompatible with the main underlying assumption of LES.
\end{abstract}

\maketitle


\section{Introduction}

The complexity of turbulence is due to a wide range of nonlinearly interacting scales. The numerical simulation of a turbulent flow, in most practical applications, cannot take into account the full range of scales, due to limitations in computational resources. The principle of Large Eddy Simulation (LES) is that only the large scales are computed directly. The influence of the scales smaller than a given scale, associated to the grid-mesh of the simulation, are modeled as a function of the resolved scales. To develop consistent subgrid scale (SGS) models, criteria are needed, based on physical or mathemathical principles and sometimes on the numerical stability of the closed set of equations. It is important that these criteria are well defined and generally accepted. The last two decades have seen the emergence of a large number of new subgrid models, and it is the authors' opinion that the turbulence community must devote more efforts in developing consensual criteria than in increasing the number of models. The purpose of the present work is to investigate one possible criterion, which is the time-reversibility of a subgrid model, when the orientation of the velocity is inversed, {\it i.e.}, under the transformation $\bm u \rightarrow -\bm u$. We report the results of Direct Numerical Simulations (DNS) in which the velocity is reversed in all, or part, of the scales of the flow. These results are then compared to results from LES in which the velocity is reversed, to assess the quality of the predictions of the models and to check whether time-reversibility is a valid criterion to assess subgrid models.

\section{Theoretical and practical considerations}

In the absence of viscosity, the dynamics of the  Navier-Stokes equations (which reduce to the Euler equations), are invariant under the simultaneous  transformation $\bm u \rightarrow -\bm u$, $t \rightarrow -t$. This means that if at an instant $t$ the velocity is reversed, the flow will evolve backwards in time until the initial condition is reached. On the level of energy transfer between scales, this property implies that in the inviscid case the direction of the energy transfer reverses when the velocity is reversed (or when time is reversed). This can be understood since the nonlinear interactions, which govern the cascade of energy between scales, are associated with triple velocity correlations. The sign of these triple products is changed when the velocity is reversed, so that the nonlinear energy transfer proceeds in the opposite direction. This symmetry is broken as soon as viscous dissipation is introduced since the viscous term of the Navier-Stokes equations does not share this symmetry. Indeed, the conversion of kinetic energy to heat through the action of viscous stresses is an irreversible process within the macroscopic (continuum) description of turbulence. Pinning down to what extent this symmetry property of the Euler equations is retained in Navier-Stokes turbulence is an interesting academic question in its own right. 

There is, however, also a practical reason to investigate this property. Even though the probability of a complete velocity reversal in a real-life flow is small, it will occur locally in time and space in various applications. Indeed in the presence of external forces which generate large-scale structures, quasi-two dimensionalisation can be observed in which the backscatter can exceed the forward flux of energy. Typical examples are thermal convection \cite{Ahler2009}, turbulence in the wake of a cylinder \cite{Meneveau}, the turbulent boundary layer \cite{Friedrich-Energy-Transfer, Fang-JOT2011} and quasi-two-dimensional flows \cite{Sommeria1986}. In these cases, the large scales can be regarded as partly reversed non-equilibrium states, which constitutes a challenge for SGS models \cite{Meneveau2000}. Some mixed models are considered good choices in numerical tests \cite{Vreman1997} for these particular flows, but the reasons are not entirely clear. In particular in these complex flow geometries it is hard to disentangle the influence of the backscatter from the influence of other flow-properties. For this reason it seems helpful to carefully assess the influence of the time-reversibility of Navier-Stokes turbulence in the academically most simple setting, isotropic turbulence. We hope that this study will thereby contribute to the understanding and evaluation of subgrid scale models for large eddy simulation.

If we consider LES and we reverse the resolved velocity, it is not known what the subgrid model is supposed to do. For convenience we will limit our discussion to the most widely used class of models, based on the concept of eddy-viscosity\cite{Smagorinsky}.  The  (scalar) eddy-viscosity model expresses the subgrid stress, 
\begin{equation}
\tau_{ij}^<=(u_iu_j)^<-u_i^<u_j^<
\end{equation}
as a function of the resolved scales ($a^<$ denoting a filtered quantity), by assuming that $\tau_{ij}^<$ is aligned with the resolved strain-rate tensor,
\begin{equation}
S_{ij}^<=\frac{1}{2}\left(\frac{\partial u_i^<}{\partial x_j}+\frac{\partial u_j^<}{\partial x_i}\right).
\end{equation}
The eddy-viscosity assumption is then given by
\begin{equation}
\tau_{ij}^<-\frac{1}{3}\tau_{mm}^<\delta_{ij}=-2\nu_t S_{ij}^<,
\end{equation}
with $\nu_t$ the eddy-viscosity. Note that here we only consider the
effect of a filter but not the discretization error. Although
it may not be the only choice (see for example Carati \textit{et al.}
\cite{Carati2001} who introduced a ``subgrid scale stress'' which
includes the error of discretization), it is the choice made in most
investigations of subgrid scale models. 

\begin{figure}
\centering
\includegraphics [width=0.5\textwidth] {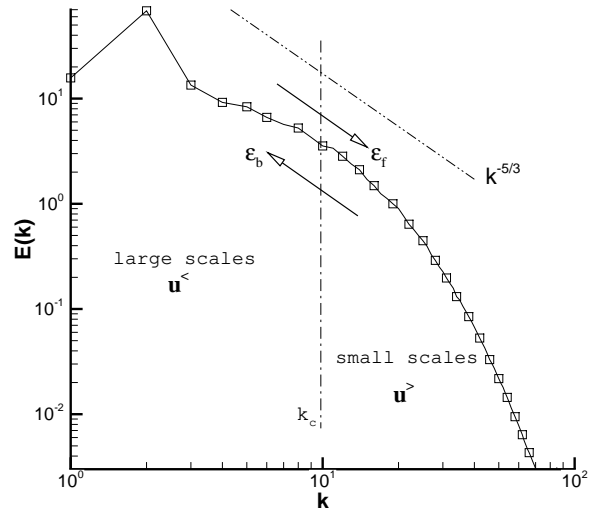}
\caption{Energy spectrum at the time of reversal. The vertical line is the location of filter. In the case of LES, The large scales are resolved and the small scales as well as the spectral fluxes have to be modeled.}
\label{fig:Ek_0.81}
\end{figure}

For some models, the reversal of the velocity leads to a reversal of the subgrid stress tensor, for others it does not.  Indeed, the dynamic procedure \cite{Germano} leads to subgrid models that are time-reversible in the sense that the dynamics evolve backwards in time after a transformation $\bm u \rightarrow -\bm u$ (see also reference \cite{Carati2001}). Another reversible model is the CZZS model \cite{CZZS} by Cui \textit{et al.} as well as its recently proposed extension \cite{IVI}. In the present work the simplified formulation of the CZZS model is used as an example of a time-reversible model. In this model the eddy-viscosity is given by
\begin{equation}\label{eqCZZS}
\nu_t = -\frac{1}{8}\frac{D_{lll}^<}{D_{ll}^<}\Delta,
\end{equation}
 where $D_{ll}^<=\langle(u_1^<(x_1+\Delta) - u_1^<(x_1))^2\rangle$ is the second-order longitudinal structure function of the filtered velocity, $D_{lll}^<$ is the third-order longitudinal structure function, $\Delta$ is the filter size and $\langle~\rangle$ indicates an ensemble average which is in practice often treated as an average in the homogeneous directions.   
This model is time-reversible since the third-order structure function changes sign when $\bm{u}^< \rightarrow -\bm u^<$. For the Smagorinsky \cite{Smagorinsky} model this is not the case. For
this model, the eddy-viscosity is given by
\begin{equation}\label{eqSmag}
\nu_t=\left(C_s \Delta\right)^2\sqrt{S_{ij}^<S_{ij}^<}.
\end{equation}
The eddy-viscosity in Eq. (\ref{eqSmag}) can not become negative, so that the net flux of energy to the subgrid scales is always positive. This flux is defined as
\begin{equation}
\Pi=\epsilon_f-\epsilon_b=-\left<S_{ij}^<\tau_{ij}^<\right> 
\end{equation}
  (see Fig. \ref{fig:Ek_0.81} for a graphic representation of the fluxes $\epsilon_f$ and $\epsilon_b$). In the Smagorinsky model the flux of energy $\epsilon_f$ from the large to the small scales is thus always larger or equal to the backscatter $\epsilon_b$. 

The time-reversibility property of models, such as for example the
dynamic model, is sometimes seen as a weakness. One reason for that is that subgrid-scale models are generally supposed to dissipate the energy flux towards the small scales (see {\it e.g.} reference \cite{Pumir2003} for a theoretical discussion on this subject). Another reason is that these models become more easily (numerically) unstable.  However it is well known that the backscatter of energy $\epsilon_b$, to the resolved scales is a physical property, which should be taken into account in a correct model of the subgrid dynamics (see {\it e.g.} reference \cite{Berto1985,Schumann1995}). Indeed the backward energy flux $\epsilon_b$ is not necessarily constrained to be inferior to the forward flux $\epsilon_f$, and a negative energy flux should therefore not {\it a priori} be excluded by a model.  

\section{Results for reversed turbulence from Direct Numerical Simulations }

\subsection{How respond the large scales when the velocity vector is reversed?}

In the present section we consider by direct numerical simulation the dynamics of subgrid and resolved scale energy after transformation $\bm u \rightarrow -\bm u$. We will define a resolved velocity field, $\bm u^<$ and a subgrid velocity  $\bm u^>$ (see Fig. \ref{fig:Ek_0.81}). In the DNS both velocities are computed and distinction between the two velocities is made by introducing a cut-off wavenumber $k_c$, corresponding to the use of a sharp spherical low-pass filter in Fourier space. Other filters could also be considered, such as Gaussian filters. We expect that the remainder of the present analysis will still hold qualitatively, but a separate investigation of the influence of the filter-type is outside the scope of the present article. Extensive investigations on the influence of the filter-type on energy transfer in isotropic turbulence can be found in references \cite{Domaradzki2007,Eyink2009,Aluie2009}. In these studies it is shown that the qualitative features (and in particular the locality) of the subgrid-scale flux  are not changed when considering smooth or sharp filters, as long as the smoothing is not too gentle. In the following we will focus on sharp spectral filters only.
The main focus of the present work is on the case in which all scales of a freely decaying turbulent flow are reversed at a given time $\tau_R$. This case will be denoted by RR, and will be compared to a freely decaying unmodified flow, denoted by NN. 

Simulations are carried out using a standard pseudo-spectral solver and a fourth order Runge-Kutta time-integration scheme, with a semi-implicit treatment of the viscous term.  The computational domain has $256^3$ grid-points. 
All cases simulate a freely decaying isotropic turbulence, starting from the same random initial field \cite{Rogallo1981}, with a spectral energy distribution similar to the measured spectrum in the experimental work of Comte-Bellot and Corrsin \cite{ComteBellot1966}. The location of the filter is illustrated in Fig. \ref{fig:Ek_0.81}, in which the energy spectrum at the time of reversal is also shown.

\begin{figure}
\centering
\includegraphics [width=0.5\textwidth]{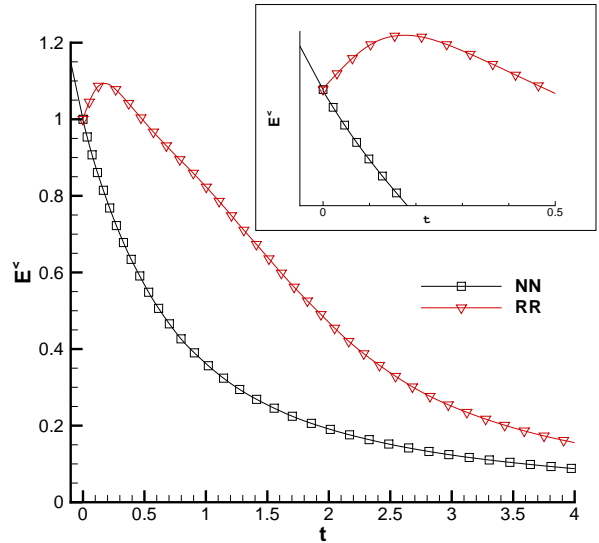}
\caption{Evolution of grid-scale energy for freely evolving turbulence compared to the evolution when the velocity vector is reversed in every point in space. In the inset we show a zoom of the behavior around the time of reversal.}
\label{fig:E_GS}
\end{figure}

 The evolution of the grid-scale and subgrid-scale energies, $E^<$ and $E^>$ respectively, is given by
\begin{eqnarray}\label{eq:DeDt}
\frac{dE^<}{dt}=-\epsilon^{<}-\Pi\\ 
\frac{dE^>}{dt}=-\epsilon^{>}+\Pi \label{eq:DeDt-2}
\end{eqnarray}
In these equations $\epsilon^{<}$ and $\epsilon^{>}$ are the grid-scale and subgrid-scale dissipation rates. At high Reynolds numbers $\epsilon^{<}$ is small compared to the energy-flux $\Pi$, if $k_c$ is chosen in the energy containing or inertial range. 
The evolution of grid-scale energy is shown in
Fig. \ref{fig:E_GS}. The time $t$ is normalized as $t\rightarrow (t-\tau_R)/\mathcal T$, with $\tau_R$ the time of reversal and $\mathcal T$ the turnover-time at the time of reversal, defined as $\sqrt{3/2}(E/\epsilon)$, with $E$ the kinetic energy and $\epsilon$ the viscous dissipation rate. The energy is normalized by $E^<(\tau_R)$, the resolved energy at the time of reversal. In the following $\epsilon^{<}$, $\Pi$, $\epsilon^{>}$ and $\epsilon$ are all normalized by $\epsilon^{<}(\tau_R)$.

The evolution of the energy of the reversed case changes radically with respect to the unmodified flow. A closer look at small times, as displayed in the inset in Fig. \ref{fig:E_GS}, shows that the energy of the grid scales in the reversed case increases, following approximately the relation $E^<(\tau_R+t)= E^<(\tau_R-t)$, as would be expected from reversible Euler dynamics, until, at later times, it starts to decay again. The increase of energy corresponds to the energy which flows back from the small scales to the large scales since the energy-cascade is reversed. The main reason that the energy level does not reach its initial value is that some of the energy of the flow has been dissipated and this process is irreversible. However, for the large scales, at which the direct influence of the viscosity is weak, the flow behaves as if it were governed by the Euler equations. A quantification of the energy flux $\Pi$ and dissipation rates $\epsilon^{<}$ and $\epsilon^{>}$ in equation (\ref{eq:DeDt}) and (\ref{eq:DeDt-2}) will be given in the following section.

This time-reversibility property of the large scales of a turbulent flow is the most important observation of this investigation: the increase of energy in the grid scales is a genuine physical effect described by the Navier-Stokes equations. The fact that a model does allow an increase of energy, such as the dynamic model, is therefore not a criterion to reject it. The opposite question: {\it should a model possess the property of reversibility to be a sound subgrid scale model?} is a different question and we will now focus on that.

\subsection{The influence of the subgrid scales}\label{secrnrz}

Evidently in a  Large Eddy Simulation we do not know the small scales. The relative insensitivity of the resolved scales on a change of the subgrid scales is the basic assumption of Large Eddy Simulation. In the context of the time-reversibility property of turbulence, we test in this section several cases in which the resolved scales and the subgrid scales are modified independently. In addition to the normal and reversed cases discussed in the previous section, we investigate here 4 different cases. In two of them the large scales are reversed but the small scales are either left unmodified or set to zero. In the other two cases the large scales are not reversed, but the small scales are again either left unmodified or set to zero. The 6 different cases which are considered are summarized in Table \ref{tab:dns-param}. 
 Note that the RN and RZ (and NR and NZ) cases are straightforwardly defined using a sharp cut-off filter in Fourier-space. The extension to smooth filters would probably raise further questions, since smooth filters can be inverted. This extension is considered as outside the scope of the present investigation in which a sharp filter is used.

\begin{table}
\begin{center}
\caption{Overview of the DNS cases. The letter N denotes normal, R reversed and Z zero.}
\vspace{0.5cm}
\begin{tabular}{cl}
\hline
\hline
DNS cases & velocity\\
\hline
NN& $+\bm u^< + \bm u^>$ \\
RR& $-\bm u^< - \bm u^>$ \\
RN& $-\bm u^< + \bm u^>$ \\
RZ& $-\bm u^< + 0$ \\
NR& $+\bm u^< - \bm u^>$\\
NZ& $+\bm u^< +0$   \\
\hline
\hline
\end{tabular}
\label{tab:dns-param}
\end{center}
\end{table}

\begin{figure}
\centering
{\includegraphics [width=0.5\textwidth]{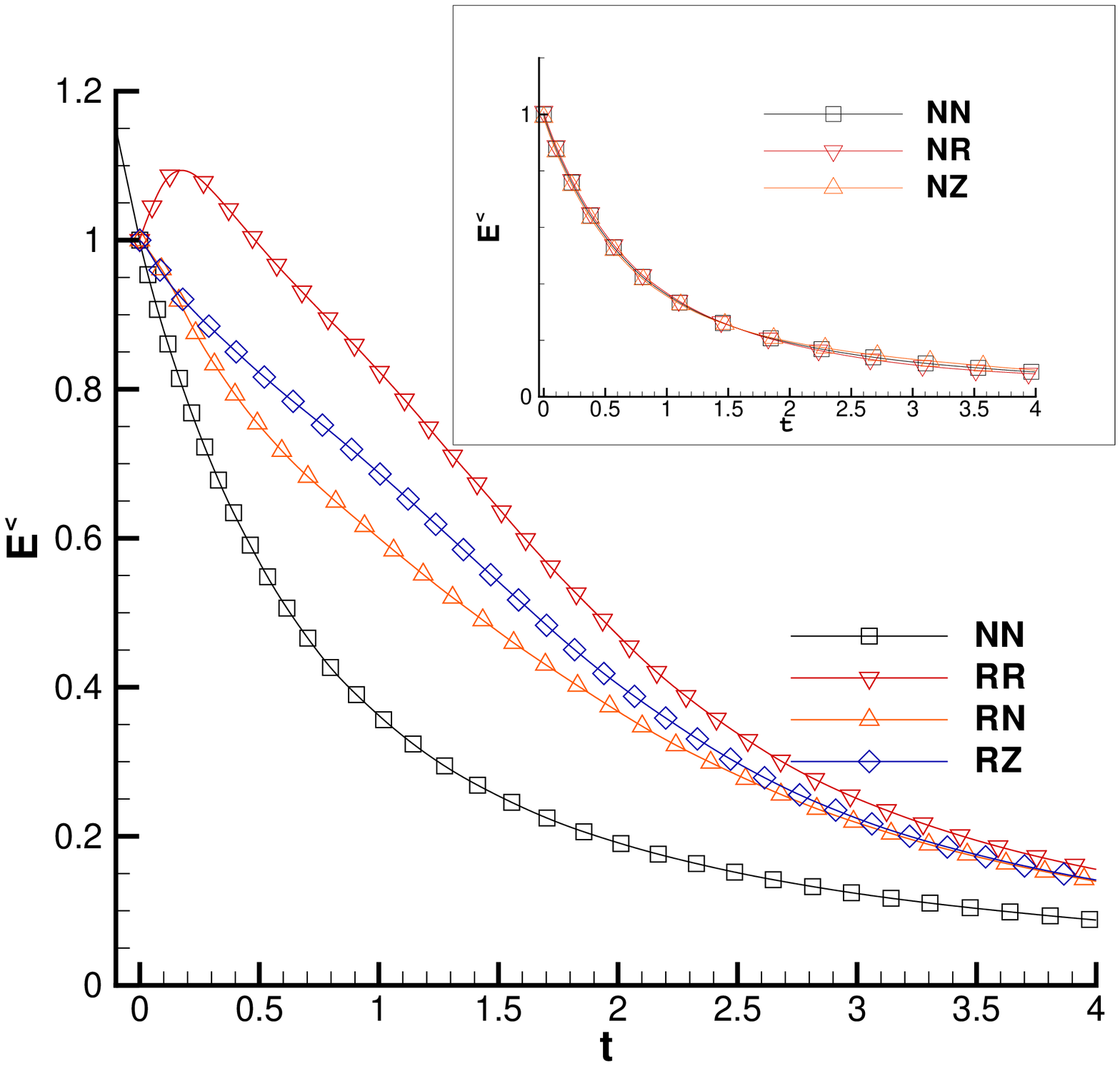}}~
{\includegraphics [width=0.5\textwidth] {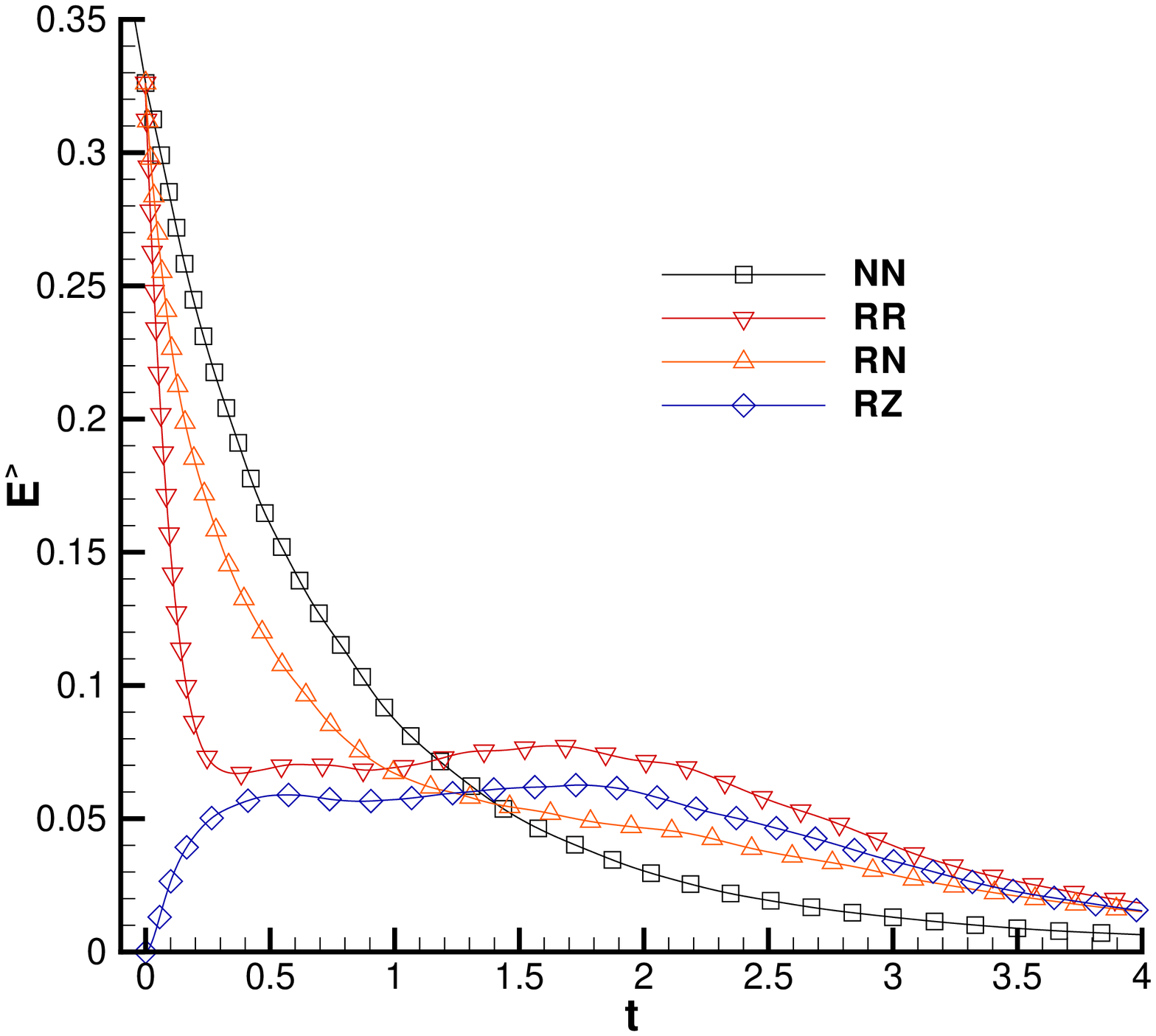}}
\caption{(a) Evolution of grid-scale energy for freely evolving turbulence compared to the evolution when the velocity vector of the large scales is reversed, but the velocity of the small scales is either unmodified (RN) or set to zero (RZ).  In the inset we show the behavior where the large scales are left unchanged, but the small scales are modified. (b) Evolution of subgrid-scale energy.}
\label{fig:E_GS-2}
\end{figure}

\begin{figure}
\centering
\includegraphics [width=0.5\textwidth]{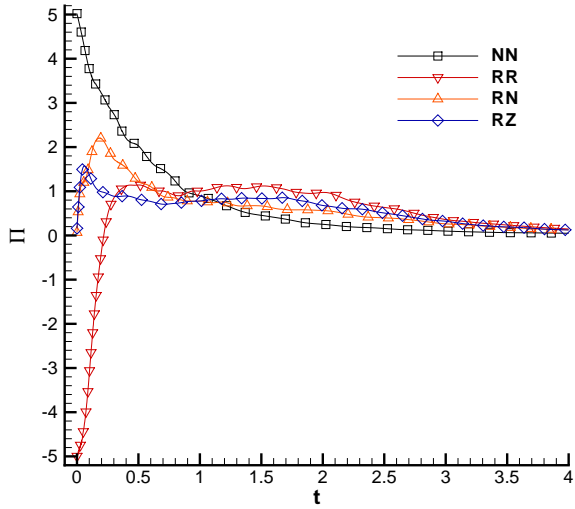}
\caption{Evolution of the energy flux $\Pi$ from large to small scales for the different runs. See table 1 for definitions.}
\label{fig:Pi}
\end{figure}

\begin{figure}
\centering
{\includegraphics [width=0.5\textwidth] {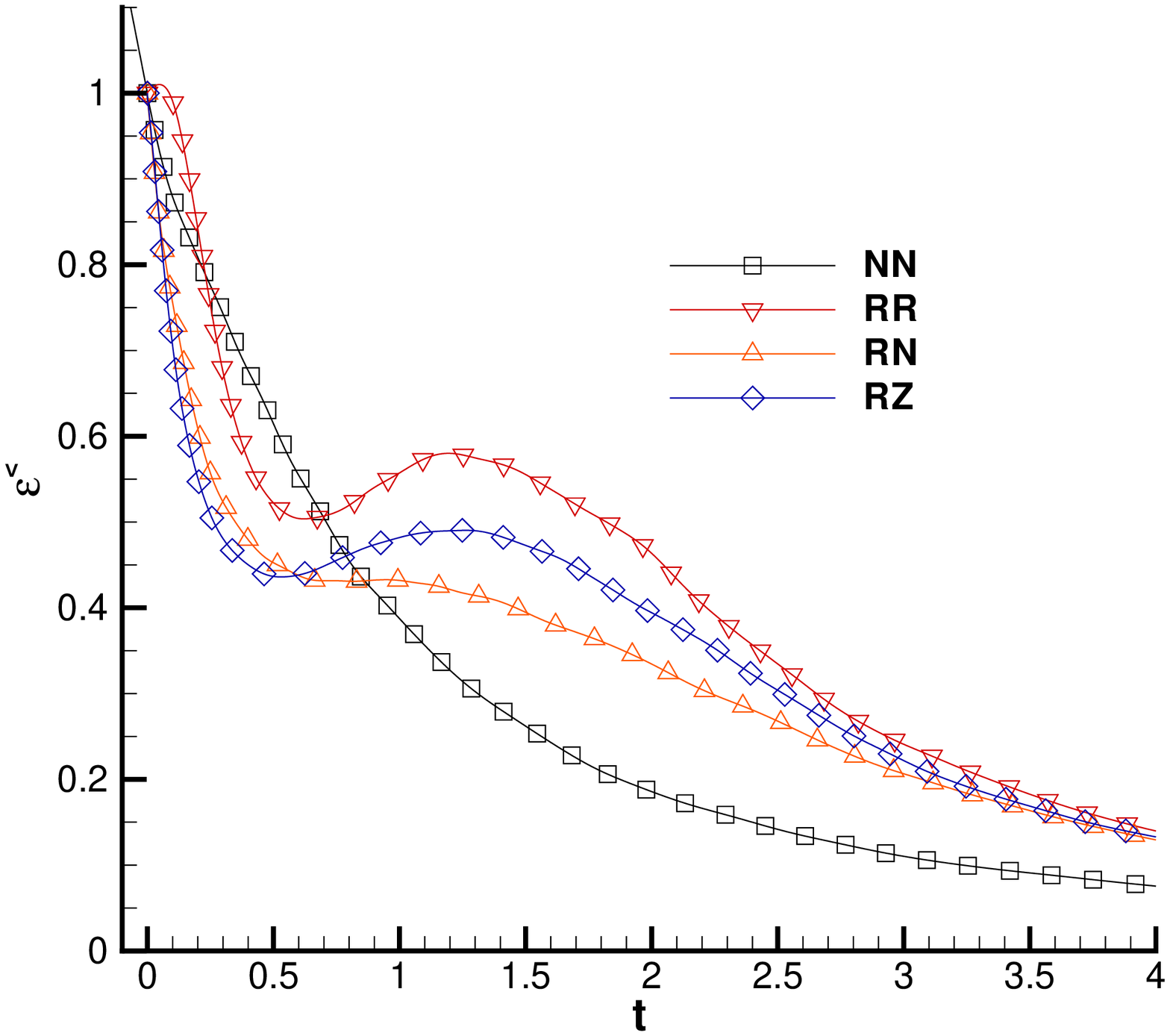}}~
{\includegraphics [width=0.5\textwidth] {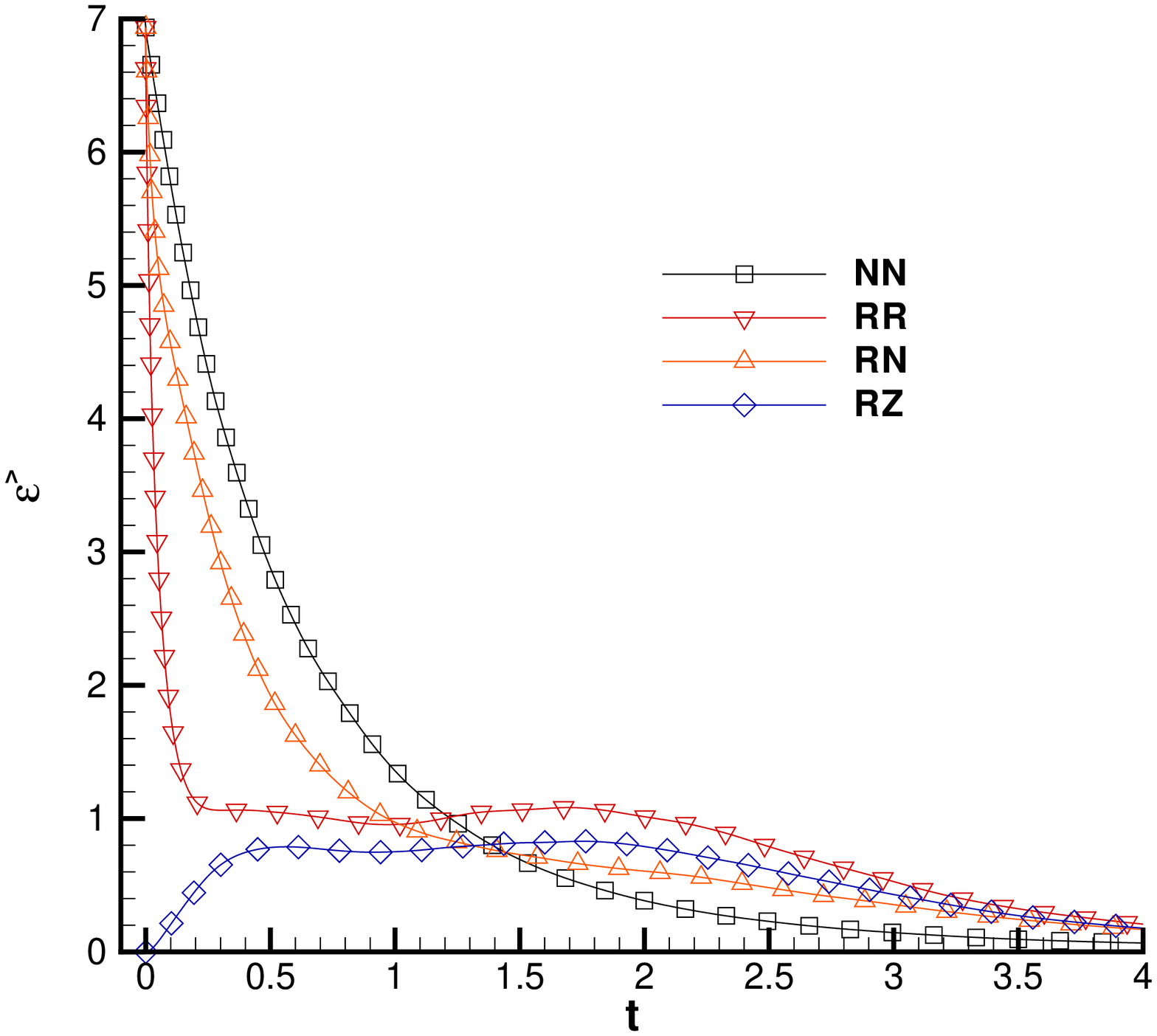}}
\caption{Evolution of (a) grid-scale dissipation and (b) subgrid dissipation for the different runs. See table 1 for definitions.}
\label{fig:eps}
\end{figure}

\begin{figure}
\centering
{\includegraphics [width=0.5\textwidth]{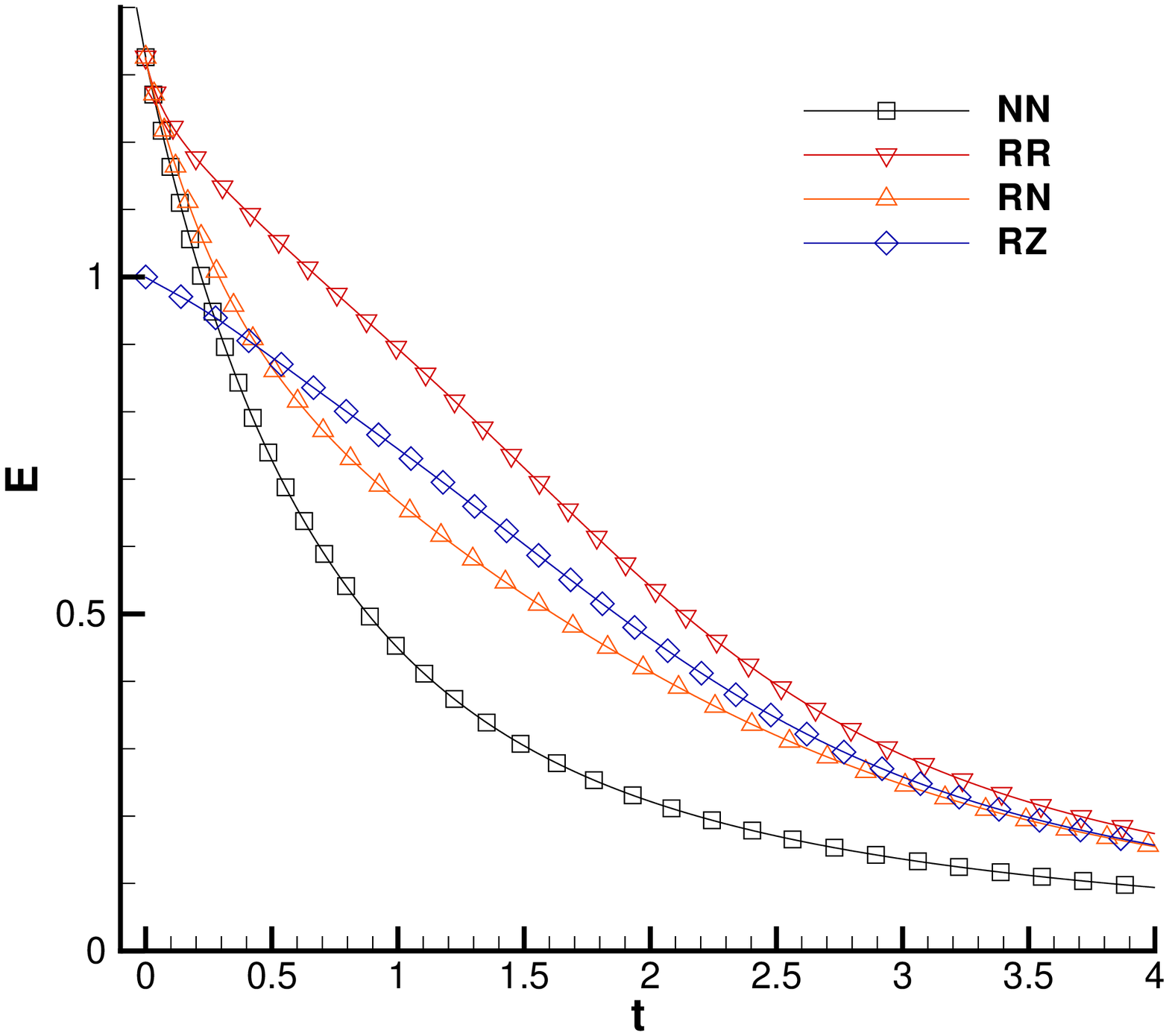}}~
{\includegraphics [width=0.5\textwidth]{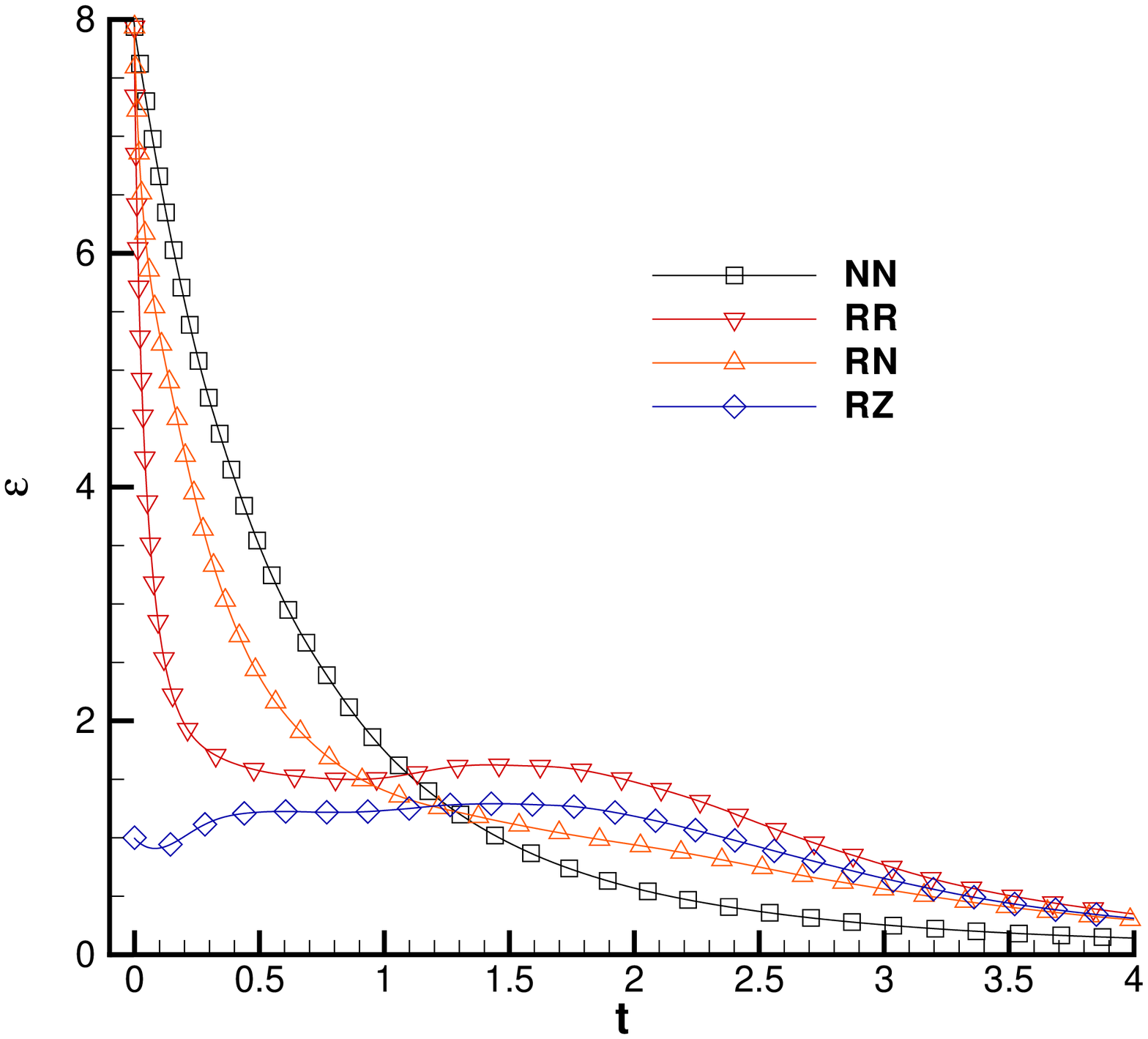}}
\caption{Evolution of the total energy and dissipation. See table 1 for definitions.}
\label{fig:Tot}
\end{figure}

In Figure \ref{fig:E_GS-2} (a) the behavior of the large scale energy is shown for the different cases. In contrast to the previous results in which all scales were reversed (the RR case), here none of the different cases displays a significant increase of the energy. However, in the cases RN, RZ in which the large scales are reversed, the energy  decay is slowed down compared to the unmodified case.  The RZ case decays more slowly than the RN case. Indeed, small scales can act as a non-local eddy-viscosity on the larger scales, an effect which is absent when these scales are set to zero. For the NR and NZ cases, in which the large scales are unmodified, the decay is not significantly altered by a reversal of the small scales. This last test can be considered a validation of one of the main assumptions of LES, {\it i.e.}, the fact that the resolved scales (in a normal, non-reversed flow) are relatively insensitive to the subgrid scale dynamics. We will not focus more on these two cases in the following.

The evolution of subgrid scale energy for the cases NN, RR, RN and RZ are shown in Fig. \ref{fig:E_GS-2}(b). The differences exist mainly in the range $0<t<0.5$. The energy of the RR and RN cases decrease very fast after reversal, since in addition to the viscous dissipation, which acts in all cases, the reversal of the grid-scale velocity leads to a reduction of the energy-input to the subgrid scale part. After some time, when the triple correlations around the cut-off are restored to transfer in the normal direction, this energy flows back into the subgrid scales, leading to a temporary energy increase for the RR and RZ cases. 

The energy flux $\Pi$ from the resolved scales to the subgrid scales is shown in Figure \ref{fig:Pi}. As expected, the energy flux at the time of reversal reverses for the RR case. For the RN and RZ cases the energy flux is strongly reduced by the reversal, but rapidly the flux is reestablished.
The dissipation rates $\epsilon^{<}$ and $\epsilon^{>}$ are shown in Figure \ref{fig:eps}. In this figure it is observed that at the time of reversal the subgrid-scale dissipation is dominant, as is expected for moderate and high Reynolds numbers. 

For completeness, we show in Figure \ref{fig:Tot}(a) the total energy. Due to the normalization by the grid-scale energy (which allowed a better comparison for the grid-scale dynamics in Figure \ref{fig:E_GS-2}) only the RZ case has unity energy at the time of reversal, since the subgrid energy is zero in this case, whereas the other cases have a higher energy. The total dissipation [Fig. \ref{fig:Tot}(b)] behaves qualitatively similar to both the subgrid dissipation and the subgrid energy.

\subsection{The influence of the Reynolds number on the time-reversibility}\label{sec:vis}

\begin{figure}
\centering
\includegraphics [width=0.5\textwidth] {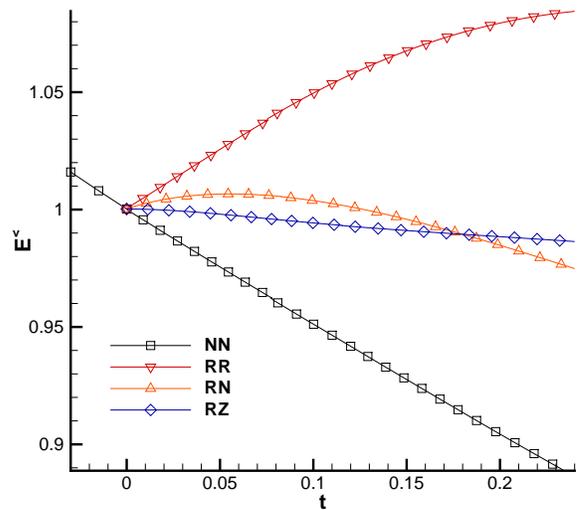}
\caption{Evolution of the grid-scale energy for the four cases (NN,RR,RN,RZ) replacing the viscous term of the Navier-Stokes equation by a $4^{\textrm{th}}$ order hyperviscous term.}
\label{fig:E_GS-HV}
\end{figure}

In the present simulations the Reynolds number is moderate and from Figure \ref{fig:eps} it can be concluded that the contribution of the grid-scale dissipation is non-negligible and of the order of $12\%$ of the total dissipation at the time of reversal. To evaluate the dependence of the Reynolds number, we would like to diminish the direct effect of the viscous dissipation on the resolved scales. We therefore calculated another set of flows replacing the viscous term by a fourth order hyperviscous draining term \cite{Borue1998,Cerutti2000}. Such a hyperviscous term concentrates the influence of the viscosity to a small range of wavenumbers so that its direct influence on the large scales is reduced. In equation (\ref{eq:DeDt}) this means that $\epsilon^{<}$ becomes very small compared to the other terms in the equations. In the present case $\epsilon^<$ is $0.15\%$ of the total dissipation at the time of reversal. 

The results of the simulations are shown in Figure \ref{fig:E_GS-HV} for short time after the reversal. It is observed that the evolution of the resolved kinetic energy at small times is qualitatively similar. The RR case shows a complete reversal at small times. The RZ case shows a slow-down of the energy decay as was observed in the normal viscous case. The only qualitative difference is observed for the RN case which now shows a slight increase of the energy, an effect which is apparently sensitive to the small amount of dissipation which was present in the viscous run. Let us digress a little and give the physical explanation for this increase. The energy transfer, analyzed in the Fourier-domain, is governed by triple products of velocity modes of the form $\left< \hat {\bm u}(\bm k)\hat {\bm u}(\bm p)\hat {\bm u}(\bm q)\right>$ with $\bm k, \bm p, \bm q$ wavevectors that can form triangles. The triple moments that are responsible for the transfer across the cut-off can be divided into two classes. One class consists of triangles with two legs of the triad shorter then $k_c$ and one longer, the other class consists of triangles with two legs of the triad longer then $k_c$ and one shorter (see also Figure 1 in reference \cite{KraichnanDIA}). In the RN case the first class of triple products will remain unchanged, since two of the three velocity modes change sign so that the triple product does not change sign. The other class will change sign since only one of the three velocity modes changes sign. The balance between the two classes and their relative contributions to the forward and backward energy fluxes will now determine whether the resulting flux is positive or negative. In the present case, apparently, the resulting flux is slightly negative, but already a small amount of viscosity is enough to prevent this reversed flux from being visible. At a later time the RN case decays faster then the RZ case, as in the viscous runs, due to the eddy-viscous effect of the small scales on the large scales which was already mentioned in the previous section.

Without focusing on the details, we only want to stress here that the precise form and location of the viscous dissipation do not qualitatively influence the behavior of the RR, and RZ cases. The RN case changes and a small increase of energy is observed.

\subsection{Assessment of the time-reversibility criterion}\label{sec:Assess}

The different behavior observed for the resolved scales when reversing the velocity in a part of the scales can be interpreted in two different ways. The first would be to point out the weakness of Large Eddy Simulation, since apparently the large scales are not independent on the details of the small scales. However this interpretation would be disingenuous, since the concept of Large Eddy Simulation in three-dimensional turbulence is intimately linked to the concept of a forward energy cascade. We would therefore prefer to point out the weakness of the criterion of time-reversibility to assess subgrid scale models. Indeed, a model should not be rejected because it is time-reversible, since at short times the dynamics of the large scales of Navier-Stokes turbulence can be reversible. However, a model which does not display this property should not be rejected either, since even in cases in which the large scales are reversed, the energy of the large scales might not increase if the energy in the subgrid scales is not reversed as is observed here in the RZ and (the viscous) RN case. With this in mind we will evaluate in section \ref{sec:LES} how different subgrid scale models behave when the velocity is reversed, but without judging on the validity of the models, which should be scrutinized using additional, less equivocal criteria.\\

\section{Results for reversed turbulence from Large Eddy Simulations }\label{sec:LES}

\begin{figure}
\centering
{\includegraphics [width=0.5\textwidth] {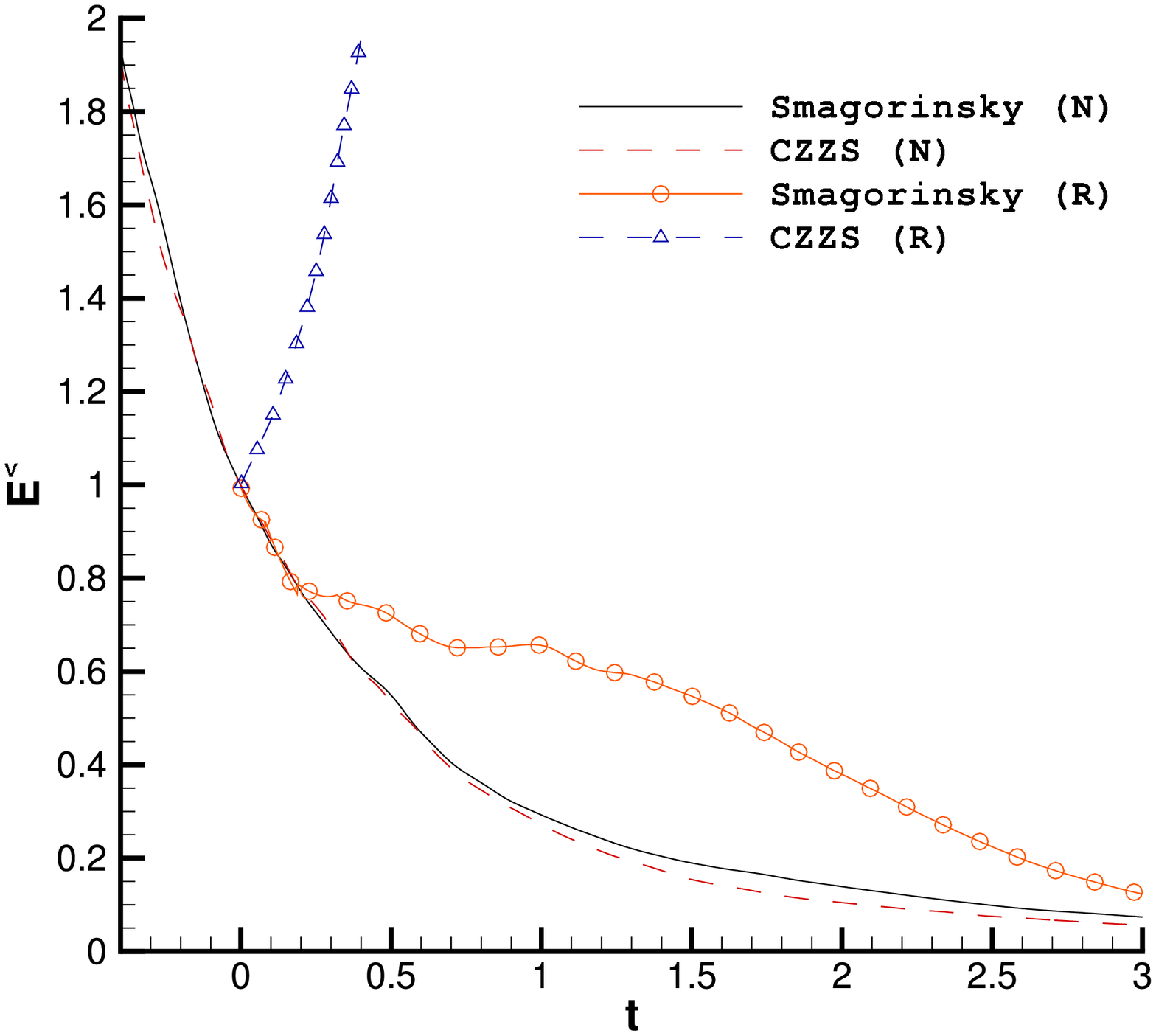}}~
{\includegraphics [width=0.5\textwidth] {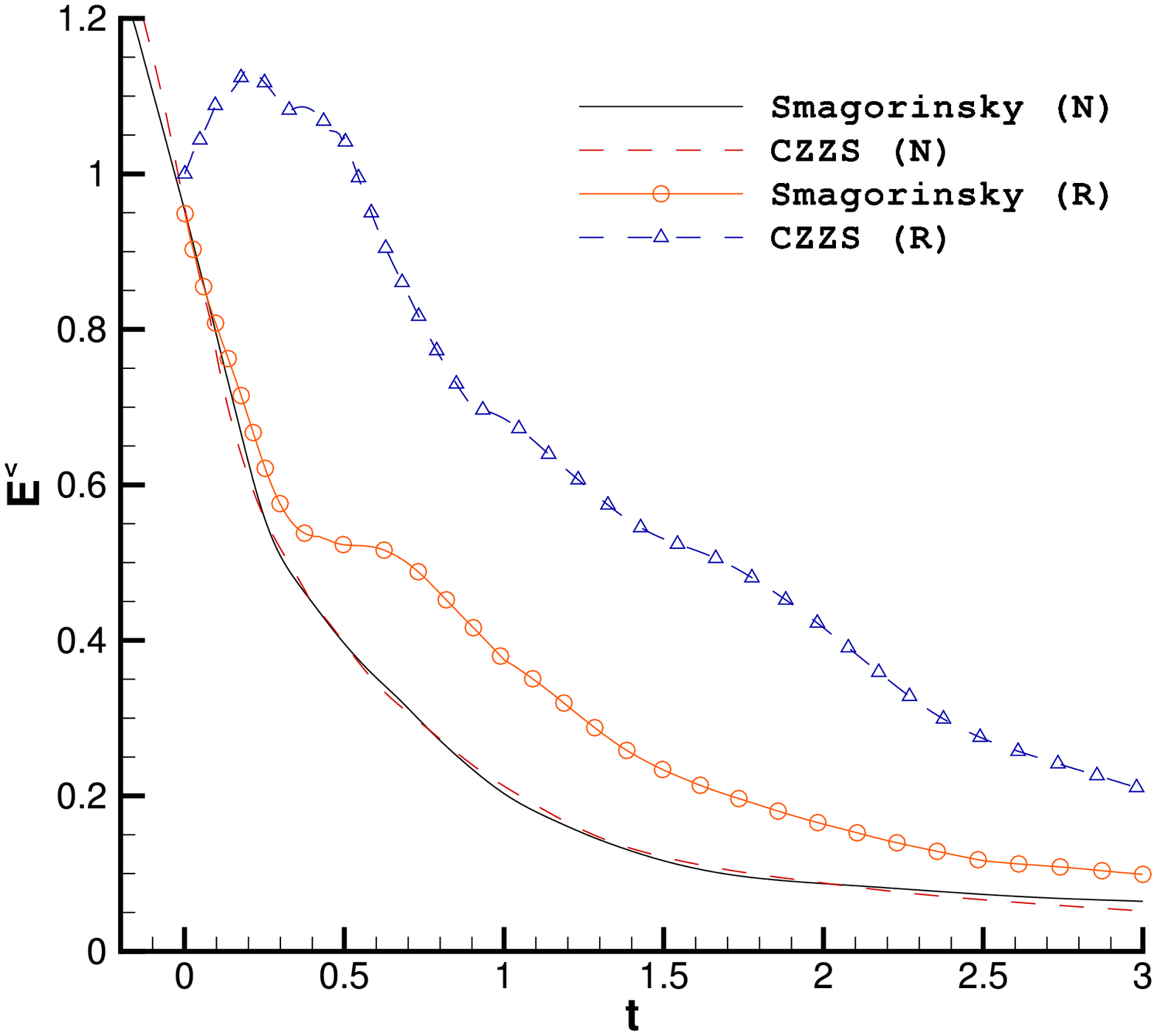}}
\caption{Evolution of grid-scale energy in the LES cases using different subgrid models. (a) without molecular viscosity (b) with molecular viscosity. N denotes normally decaying, R means reversal of the velocity.}
\label{fig:E_GS_LES}
\end{figure}

We perform in the present section the same test, reversing the large scales using different subgrid models, first the simplified CZZS model (\ref{eqCZZS}), second the Smagorinsky model, Eq. (\ref{eqSmag}) with $C_s$ fixed at $0.14$. The first model is, as mentioned before, time-reversible, the second  is not. The computational mesh has $48^3$ grid-points. As was shown by Kraichnan \cite{Kraichnan1976}, a constant (non-scale dependent) value for the eddy-viscosity is only a good approximation in the inertial range, far from the cut-off frequency. Close to the cut-off, where the role of the model is most important, the value of the eddy-viscosity strongly increases. This effect can be corrected for by adding a scale dependent cusp to the model, as was applied by Chollet and Lesieur \cite{Viscosity-spe}. This should be done in principle for {\it all} eddy-viscosity models. In the present work this cusp is introduced by modifying the eddy-viscosity to
\begin{equation}
\nu_t^*(k)=\nu_t (1+34.6 \exp(-3 k_c/k)).
\end{equation} 
Two different Reynolds number cases are considered. In the first one, the viscosity is set to zero, yielding, in the absence of $\nu_t$, the time-reversible Euler equation. The comparison of grid-scale energy is shown in Fig. \ref{fig:E_GS_LES}(a). We can observe that after reversal, the simplified CZZS model yields an increase of energy, which is similar to what was observed in the RR case in the last part, before the irreversible influence of viscosity set in. The Smagorinsky model remains decaying at the same rate as the normally decaying case for some time-steps after reversal. This phenomenon is not similar to any DNS case in the last part, and stems from the fact that reversal leaves the value of the eddy-viscosity unchanged since $S_{ij}^< S_{ij}^<$ is unchanged after reversal. For longer times, the decay rate decreases, since the direction of the resolved energy cascade changed sign. However the energy cannot increase, since $\nu_t$ can not change sign. 

In order to determine the influence of the Reynolds number on the dynamics, we considered a second decay-case, in which the molecular viscosity was not set to zero. In this case the ratio $\nu/\nu_t\approx 1$ at the time of reversal, and the results are shown in Fig. \ref{fig:E_GS_LES}(b). In this case the behavior of the reversible model is very close to the DNS result of the RR case, for both small and long times. The presence of a non-negligible amount of non-reversibility by the viscous stress, prevents the flow from developing an unlimited amount of energy. We want to stress however that the apparent success of the reversible model in the presence of non-zero viscosity in reproducing the DNS results of the last section is fortituous and depends on the Reynolds number. In other words, it might not always be easy to foresee which amount of viscosity is needed to avoid non-physical effects. At high Reynolds numbers the behavior might ressemble more the inviscid behavior shown in Fig. \ref{fig:E_GS_LES}(a). The underlying issue here is that a subgrid model needs to reproduce two distinct features of the subgrid scales. The first is to drain the energy from the large scales. The second is to dissipate this energy. The dynamic model only forefills the first task, which corresponds to the reversible interaction between the small and the large scales, but does not dissipate the energy.

\begin{figure}
\centering
{\includegraphics [width=0.5\textwidth] {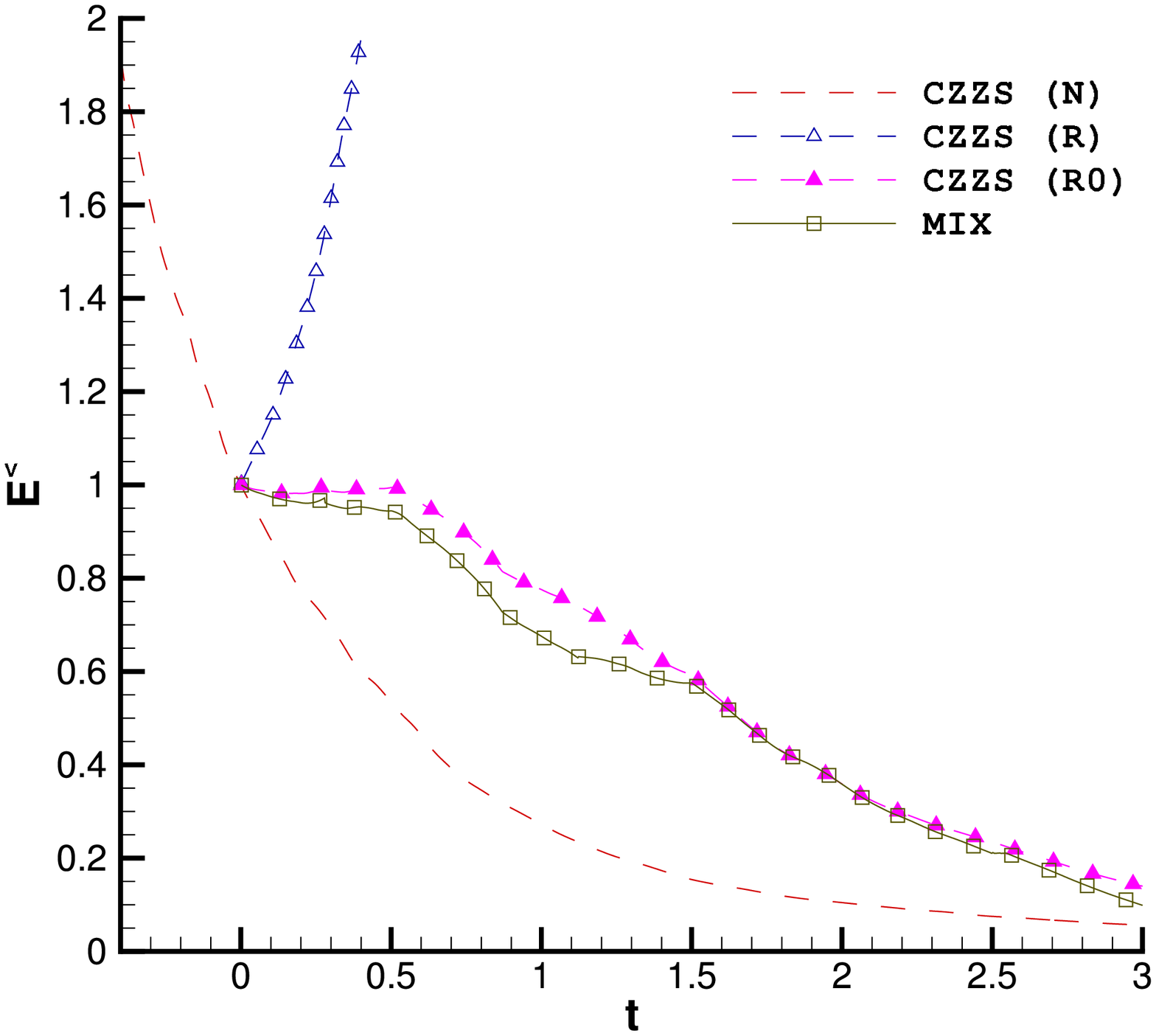}}~
{\includegraphics [width=0.5\textwidth] {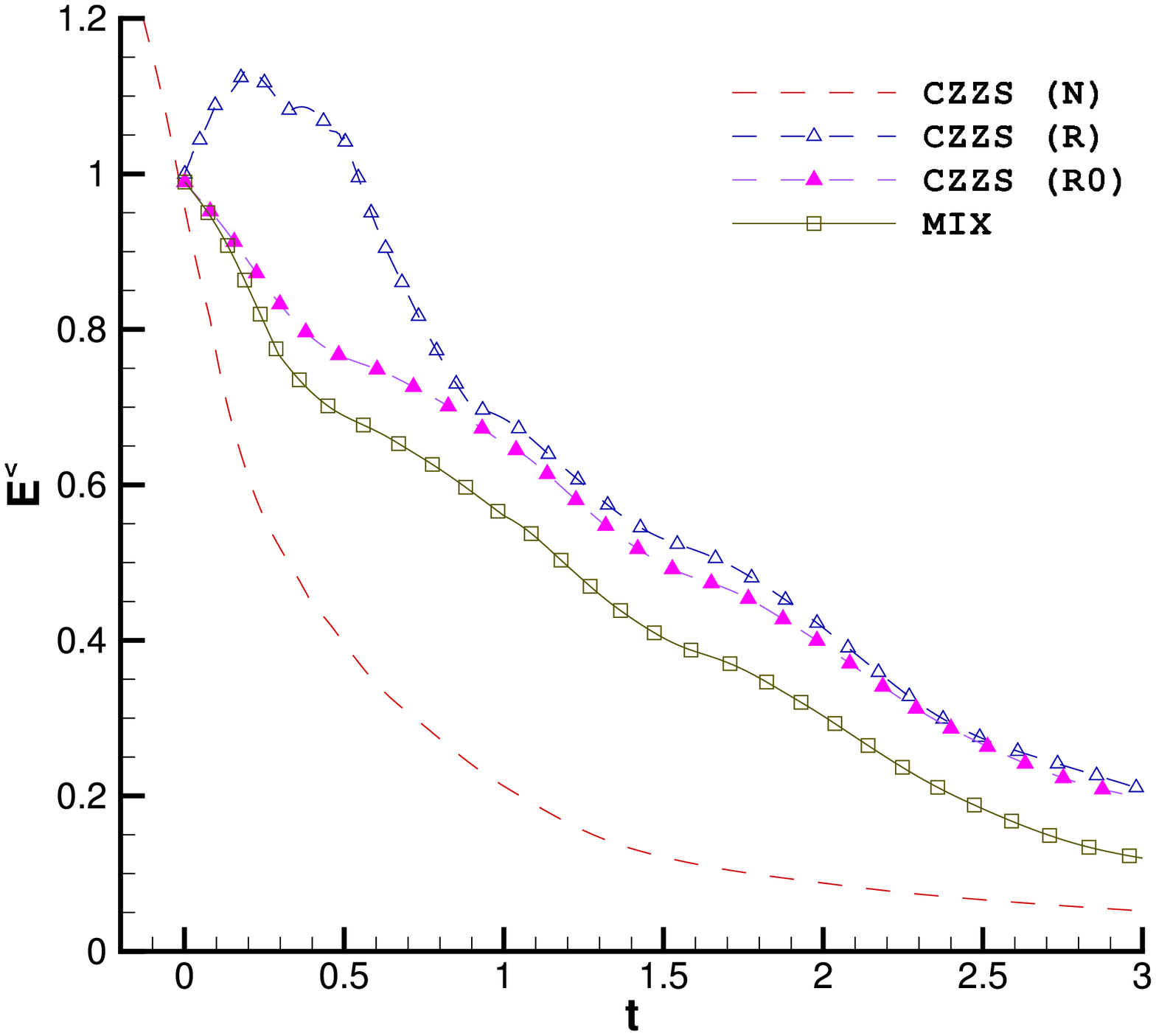}}
\caption{Evolution of grid-scale energy in the LES cases using different subgrid models. (a) without molecular viscosity (b) with molecular viscosity. N denotes normally decaying, R means reversal of the velocity, and R0 means fixing the negative viscosity as zero.}
\label{fig:E_GS_LES-2}
\end{figure}

For the sake of completeness, we also test two other models, which might be of practical use if one does not want to worry about the amount of viscosity needed to avoid a reversible model to reinject unphysical amounts of energy in the system. For the first one (model R0 in the following) we use the simplified CZZS model in which we reverse the velocity, but fix all negative viscosity as zero, \textit{i.e.}
\begin{equation}\label{model:R0}
 \nu_t = \max(-\frac{1}{8}\frac{D_{lll}^<}{D_{ll}^<}\Delta, 0).
\end{equation} 
In real practice, this clipping procedure was widely used in the time-reversible models to obtain numerical stability. For the second one (that we denote MIX in the following) we follow the strategy of defining a mixed model as suggested by Vreman \textit{et al.} \cite{Vreman1997} and
\begin{equation}\label{model:MIX}
 \nu_t = \frac{1}{2}\left(-\frac{1}{8}\frac{D_{lll}^<}{D_{ll}^<}\Delta + \left(C_s \Delta\right)^2\sqrt{S_{ij}^<S_{ij}^<} \right),
\end{equation} 
where the additional coefficient $1/2$ is used to guarantee the consistency with non-reversed turbulence. These mixed models often lead to good results in real applications \cite{Vreman1997}, but their formulation is not supported by theoretical or physical arguments.

Shown in figure \ref{fig:E_GS_LES-2} is the behavior of the models defined in expressions (\ref{model:R0}) and (\ref{model:MIX}). We observe that their behavior closely resembles the RN and RZ cases in Fig. \ref{fig:E_GS}, where grid-scale energy decay is reduced during a short time, and then decays normally. We can therefore conclude that the models (\ref{model:R0}) and (\ref{model:MIX}) represent a physical behavior, corresponding to a certain class of flows. 

As we argued in the previous section, we cannot use the present results to assess the models. The backflow of energy is not unphysical but it is a phenomenon which is not observed in all possible flows in which the resolved scales are reversed. If in a particular application one aims at the prediction of a time-reversed flow without risking an unlimited amount of backscatter, one can use one of the models given by (\ref{model:R0}) and (\ref{model:MIX}). A more sophisticated, but also more physical, procedure was proposed by Ghosal {\it et al.} \cite{Ghosal1995} by basing the flux of energy to the small scales on the subgrid scale energy, which was computed using a transport equation for the Reynolds averaged subgrid scales. This procedure implicitly assumes that the net energy flux from the resolved scales to the subgrid scales is determined by the subgrid energy. Another solution would be to introduce a cascade time, based on the energy around the cut-off, which limits the time during which the model remains time-reversible. However, all these fixes are only needed if one wants to be able to take into account the time-reversibility of the resolved scale turbulence.

\section{Conclusion}

The {\it gedanken}-experiment in which the velocity at each point in a turbulent flow is reversed can be carried out in numerical simulations, and this is what we performed in the study presented in this work. The goal of this work was not to judge particular subgrid scale models or even the whole concept of LES using the criterion of time-reversibility, but rather to judge the criterion itself. Our conclusion is then the following: the property of time-reversibility alone is not an unequivocal criterion to reject or qualify subgrid models.

We base this judgment on two observations. The first one is that at short times the energy of the resolved scales increases in a reversed flow, as would be the case for a flow governed by the (truncated) Euler equations. This observation alone could be used to argue that subgrid models should be, at least partly, reversible, and to reject subgrid models which do not possess this property. However, a second observation in the present work showed that if we reverse the large-scales but do not modify the subgrid scales or set them to zero, the large-scale energy does not necessarily increase. This second observation shows that, even if a model cannot increase the energy of the resolved scales, it still corresponds to a certain class of flows, and the model cannot be rejected. However, it can not be excluded that the invariance of a model with respect to the reversal of the velocity will not have other consequences when regarding other diagnostics.

For some practical purposes, in which the user is only interested in a model that drains a sufficient amount of energy without compromising the stability of the simulation, time-reversibility will probably continue to be regarded as a possible criterion to reject a subgrid model. We claim here, on the basis of the present results, that considering the detailed flow physics, this criterion is unsuitable. This unsuitability to use the criterion of time-reversibility to assess subgrid scale models for LES is inherent to the basic assumption of Large Eddy Simulation. If LES is to be usable at all, some {\it a priori} assumption by the user should be made about the property of the cascade of energy. This cascade is in general towards the small scales in three-dimensional turbulence, and in this case the user of LES should accept that some physics are not captured by his simulations or he should give some input about the unknown scales to the model. If this is not satisfactory in some applications, LES, in its present form, might simply not be the adequate tool to study these particular applications. More sophisticated approaches might then be needed, in which the direction of the energy flux between scales can be determined as a function of resolved flow parameters or models in which the scales are not arbitrarily divided into large and small scales.
 
\section*{Acknowledgments}
The authors acknowledge interaction with Gr\'egoire Winckelmans and Robert Rubinstein. The authors also acknowledge an anonymous reviewer of a previous manuscript, who, by asking a question about the time-reversibility of subgrid-models, triggered the current investigation. L. Shao acknowledges support from BUAA SJP 111 program (Grant No. B08009).


\end{document}